\documentclass[a4paper]{article}
\usepackage{amsmath,amssymb,epsfig,latexsym}
\newcommand{\dt}[1]{\frac{d#1}{dt}}
\newcommand{\nsum}[1]{\langle#1\rangle}

\begin{document}
\title{Dynamics of order parameters for a population of globally coupled
oscillators}
\author{Silvia De Monte$^1$ and Francesco d'Ovidio$^2$\\
\\
$^1$Institut f\"{u}r Mathematik, Strudlhofgasse 4, A-1090 Wien,
Austria\\
{\it silvia.demonte@univie.ac.at}\\
$^2$Dept. of Physics, Tech. University of Denmark, DK-2800 Lyngby,
Denmark\\
{\it dovidio@fysik.dtu.dk} 
}

\date{\today}
\maketitle

\begin{abstract}

Using an expansion in order parameters, the equation of motion for the centroid
of globally coupled oscillators with natural frequencies taken from a
distribution is obtained for the case of high coupling, low dispersion of
natural frequencies and any number of oscillators. To the first order, the
system can be approximated by a set of four equations, where the centroid is
coupled with a second macroscopic variable, which describes the dynamics of
the oscillators around their average. This gives rise to collective effects
that suggest experiments aimed at measuring the parameters of the population.

\end{abstract}

%\columns[2]
%\fontsize{12}{24}\selectfont

Populations of coupled oscillators have been widely used as a framework for
studying collective phenomena. Applications include chemical systems (e.g.,
reaction-diffusion systems in the oscillatory regime [1]), electronic
circuits [2], and biological systems (e.g., spiking activity in
neurons [3], synchronization in $\beta$-cells [4], and glycolytic
oscillations [5]).

Much of the interest has been devoted to properties arising at the population
level, especially in relation with synchronization phenomena [6]. Borrowing
from statistical mechanics, such properties may be quantified in terms of order
parameters [7], such as the mean value or centroid, for which phase diagrams
can be constructed.

In this Letter, we consider a population of $N$ Hopf normal forms with a
mean-field coupling:

\begin{equation}\label{eq:zj}
\dt{z_j}=(1+i\omega_j-|z_j|^2)z_j
+K\frac{1}{N}\sum_{l=1}^{N}(z_l-z_j),
\end{equation}

where the natural frequencies $\omega_j$ are taken from a
given distribution and $K$ denotes the coupling strength.

Previous analyses [8] of system\ (\ref{eq:zj}) (or of its phase
reduction) have been restricted mainly to properties arising in the asymptotic
regimes. In this case, an insight into the qualitative behavior of the system
has been obtained by observing the asymptotic properties of the centroid
$Z=\nsum{z_j}$ (using the notation $\nsum{f_j}=\frac{1}{N}\sum_{j=1}^{N}f_j$).

Our aim is to describe the dynamics of system (\ref{eq:zj}) by expanding
it in order parameters or \emph{macroscopic variables} (as they
shall be called from now on in order to avoid confusion between control
parameters and order parameters). The explicit deduction of their equations of
motion shows how properties emerge from the microscopic level, and
allows us to describe the dynamics of the system outside the
asymptotic regime, a relevant issue in systems
interacting with the environment or in experiments
using perturbation response techniques.

The macroscopic equations are derived through a change of coordinates from the
microscopic variables $z_j$ to macroscopic variables (like the centroid) that
are averaged over the population. The main advantage of this procedure is
that the macroscopic variables can be organized in a
hierarchy, retaining, if the coupling is strong and the natural frequency
distribution is narrow, only the lowest order terms of a series expansion.
In this region of the parameter space, the equation for the centroid has a
functional form reminiscent of Eq.\ (\ref{eq:zj}), but it is also coupled with
a second macroscopic variable that describes the dynamics of the oscillators
around their average. As a consequence, with respect to the description of the
system as a single macroscopic oscillator, a correction for the collective
oscillations amplitude appears and a critical macroscopic perturbation (i.e.,
\emph{quenching}, see later) can be identified. Analytical relations for such
quantities are given and suggest simple experiments for determining, by means
of purely macroscopic measures, the coupling strength and the variance of the
natural frequencies distribution in real systems. Our results are valid for
any  number of oscillators, and for any (narrow) distribution of natural
frequencies and are compared with those obtained by numerically integrating
the original system\ (\ref{eq:zj}).

{\it Deduction.} Let us start by writing the positions of the oscillators in
terms of their distance from the centroid: $z_j=Z+\epsilon_j$.
This expansion is useful when the coupling is strong, since in
this case the dynamics leads the system to collapse on a
configuration, the so-called locked state [9], that is peaked around
the centroid. In particular, the displacements from the
centroid in the locked state converge to zero when the coupling strength is
increased$^\dag$. Moreover, when the displacements are wider than the value at
the equilibrium, the dynamics is contracting, and thus the locked state
furnishes an upper limit to the broadness of the initial configuration
required for the approximations to hold.

Eq.\ (\ref{eq:zj}) now reads:
 
\begin{eqnarray}\label{eq:epsexp}
\dt{z_j}=&&\left(1-|Z|^2+i\omega_j\right)Z+\epsilon_j\left(i\omega_j+1-K-2|Z|^2\right)\nonumber\\
&&-Z^2\epsilon_j^{*}+o(|\epsilon_j|^2). 
\end{eqnarray}

We then consider the time evolution of the centroid, differentiating its
definition: $Z=\nsum{z_j}$ and using Eq.\ (\ref{eq:epsexp}):

$dZ/dt=d\nsum{z_j}/dt=\nsum{dz_j/dt}
=(1-|Z|^2+i\nsum{\omega_j})Z+i\nsum{\omega_j\epsilon_j}+
(1-K-2|Z|^2)\nsum{\epsilon_j}-Z^2\nsum{\epsilon_j^*}+\nsum{o(|\epsilon_j|^2)}$.

By definition of a centroid,
$\nsum{\epsilon_j}=\nsum{\epsilon_j^*}=0$, and $dZ/dt$ reduces to:

\begin{eqnarray}\label{eq:dZdt}
\dt{Z}=(1-|Z|^2+i\omega_0)Z+i\nsum{\omega_j\epsilon_j}+o\left(\nsum{|\epsilon_j|^2}\right)
\end{eqnarray}

where $\nsum{\omega_j}=\omega_0$ is the average natural frequency.
A {\it zeroth-order} expansion thus leads to the equation:

\begin{eqnarray}\label{eq:zeroth}
\dt{Z}=(1-|Z|^2+i\omega_0)\,Z.
\end{eqnarray}

This expansion has the same functional form as the
individual, uncoupled elements. It exactly describes the case of a
population of oscillators with the same natural frequency $\omega_0$ and with
$\epsilon_j=0\quad\forall j$, the last condition being fulfilled if all the
oscillators are assigned the same initial condition. In this (trivial) case,
it shows the existence of a limit cycle of radius 1 and frequency $\omega_0$,
and of an unstable focus in $Z=0$. The equation is independent of the coupling
strength, since the coincidence of the oscillators and the centroid is
maintained in time. For the description of the parameter
mismatch to be included, the term of lower order that must be then
taken into account is: $W:=\nsum{\omega_j\,\epsilon_j}$. Having in
this way defined a new macroscopic variable, we now derive its
equation of motion.

Since $dW/dt=d{\nsum{\omega_j
\epsilon_j}}/dt=\nsum{\omega_jdz_j/dt}-\nsum{\omega_j }dZ/dt$, 
using the definition of $W$ and Eq.\ (\ref{eq:dZdt}), we obtain:

\begin{eqnarray}\label{eq:dWdt}
\dt{W}=i\left(\nsum{\omega_j^2}-\omega_0^2\right)Z
+\left(1-K-2|Z|^2+i\omega_0\right)W\nonumber\\-Z^2W^*
+i\nsum{(\omega_j-\omega_0)^2\epsilon_j}+o\left(\nsum{|\epsilon_j^2|}\right).
\end{eqnarray}
 
The equation is closed with respect to $Z$ and $W$, if again the
higher order terms in the displacements are discarded, when the
dispersion of the natural frequencies is sufficiently small. Since we
are dealing with narrow distributions, the second degree term in the
frequencies will be neglected. Note that this approximation does not depend
on specific symmetries or shapes for the frequency
distribution.  Calling $\sigma^2=\nsum{\omega_j^2}-\omega_0^2$ its variance,
and combining Eq.\ (\ref{eq:dWdt}) with Eq.\ (\ref{eq:dZdt}), we get the
following description at {\it first order}:

\begin{eqnarray}\label{eq:first}
\begin{cases}
\dt{Z}=(1-|Z|^2+i\omega_0)Z+i\,W\\
\dt{W}=i \,\sigma^2 Z
+\left(1-K-2|Z|^2+i\omega_0\right)W-Z^2W^*.
\end{cases}
\end{eqnarray}

This first-order expansion shows explicitly how the centroid is coupled with
the dynamics of the oscillators around it.  This ``internal'' dynamics is
quantified in the $W$ variable, by means of which the parameter mismatch
(through the variance of natural frequencies) and the coupling strength affect
the system. These equations describe the dynamics of the centroid from initial
conditions with small $\epsilon_j$ (and thus small $W$). Since in
their derivation we have requested to have asymptotically a locked
state, we will restrict our analysis to the case in which the system
has an unstable focus surrounded by a limit cycle, which occurs if
$K>1+\sigma^2$.

Although we stop at this order, the deduction can be carried
further, and macroscopic descriptions at higher orders derived,
applying the same method: the terms that have been discarded in Eqs.\
(\ref{eq:dZdt}) and (\ref{eq:dWdt}) define new macroscopic variables,
and their equations of motion can be derived by differentiating their
definition.

{\it Analysis.} If $K>1+\sigma^2$ and $\epsilon_j$ are small, it is easy to
see that the behavior of the system is essentially described by the dynamics on
the invariant and attracting manifold on which the two macroscopic variables
are orthogonal. This condition is fulfilled, and the description exact, when
the natural frequency distribution and initial configuration are both
symmetric. Setting $Z=R \,e^{i(\phi+\omega_0 t)}$ and $W=w\,
e^{i(\theta+\omega_0 t)}$, Eqs.\ (\ref{eq:first}) reduce on
$\phi=\theta+\pi/2$ to the amplitude equations:

\begin{eqnarray}\label{eq:red}
\begin{cases}
\dt{R}=(1-R^2)\,R-w\\
 \dt{w}=\sigma^2 \, R+(1-K-R^2)\,w.
\end{cases}
\end{eqnarray}
 
System\ (\ref{eq:red}) can be analyzed with
planar methods and furnishes both an estimate for the amplitude
of the locked oscillations and a description of their transient
behavior.

Let us first find the equilibrium values for $R=|Z|$ and $w=|W|$.
Setting  $\alpha=K/2-\sqrt{(K/2)^2-\sigma^2}$ and using Eq.\
(\ref{eq:red}), it is easy to show that $W$ and $Z$ display
synchronous oscillations (the second with a phase delay of $\pi/2$)
of amplitudes:

\begin{equation}\label{eq:equi}
R_1=\sqrt{1-\alpha}, \quad w_1=\alpha\sqrt{1-\alpha}.
\end{equation}

Taking into account that in our approximation the ratio
$\sigma^2/K$ is small, this expression approximately results
in:

\begin{equation}\label{eq:zf}
R_1=1-\frac{\sigma^2}{2K},\quad w_1=\frac{\sigma^2}{K}.
\end{equation}

When these expressions are compared with the zeroth-order amplitudes $R_0=1$
and $w_0=0$, it appears that the term $\sigma^2/2K$ gives the first
order correction to the amplitude of the oscillations due to the
mismatch of the natural frequencies.
It is worth remarking that previously derived formulae for the
amplitude of $Z$ were  only implicitly related to the parameters of
the population, through a self-consistency integral [9], and in
the limit of large $N$.

Let us now address the structure of the phase space. This is
particularly relevant when the perturbation response is addressed, since it
qualitatively and quantitatively determines the features of the
transient after the system is initialized in a configuration out of
equilibrium. A fundamental example is the {\it quenching}
technique [10], which consists in damping the oscillations of the
system by displacing it onto the stable manifold of the unstable focus.
This can be experimentally observed as a
long-term vanishment of the oscillations amplitude.
While in the case of the zeroth-order expansion the stable manifold
reduces to the focus itself, in the case of the first-order expansion it
consists of a two-dimensional variety, which on the invariant plane
reduces to the stable eigenspace of the unstable focus:

\begin{equation}\label{stabmani}
w-(K-\alpha)R=0.
\end{equation}

In quenching experiments the system lies in the locked state and is
then macroscopically perturbed through a rigid displacement of the
whole population. Since only the amplitude of the collective
oscillations is measured, the critical displacement is identified by
means of the {\it quenching radius}, that indicates the amplitude of the
initial state for which the oscillations are suppressed. According to
Eq.\ (\ref{eq:red}), this radius is given by:

\begin{equation}\label{eq:rq}
R_q=\frac{\alpha^2\sqrt{1-\alpha}}{\sigma^2}\cong\frac{\sigma^2}{K^2}.
\end{equation}

Again, we notice that the expression reduces to the
zeroth-order value, that is zero, in the limit of vanishing $\sigma$ and large
$K$.

\begin{figure}[H]
\center
\epsfig{file=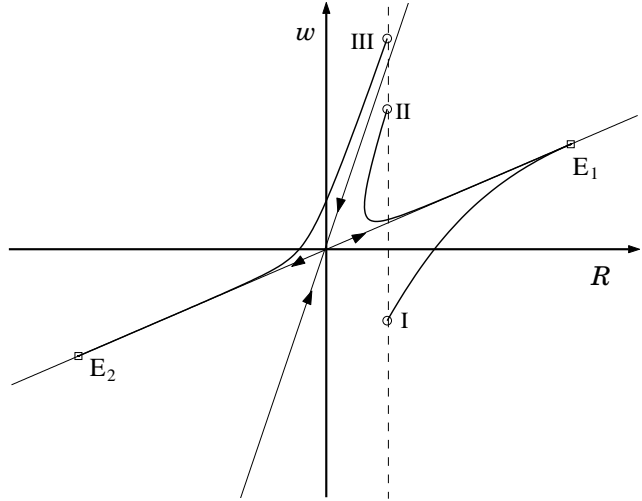,clip=,width=.7\textwidth}
\caption{Phase portrait of the reduced system Eq.\ (\ref{eq:red}). The
equilibria ($E_1$ and $E_2$) and the eigenspaces of the origin are indicated.
The system is initialized in three points (circles) giving rise to
qualitatively different trajectories.\label{fig:phase}}
\end{figure}

Moreover, as illustrated in Fig.\ \ref{fig:phase}, the structure of the phase
space allows us to predict three qualitatively different kinds of transient
amplitude increase. In one case (I), the system behaves like an ordinary
oscillator, the impulsive perturbation being followed by a monotonic increase
in amplitude. In the second case (II), the amplitude first decreases, while the
centroid is approaching the origin, and then increases again, leading at the
same asymptotic solution as in the previous case. Finally (III), if the
initial configuration width is large enough, it will first reduce to zero and
then increase again, but the asymptotic solution will be in phase opposition
with respect to the initial oscillations.

\begin{figure}[H]
\center
\epsfig{file=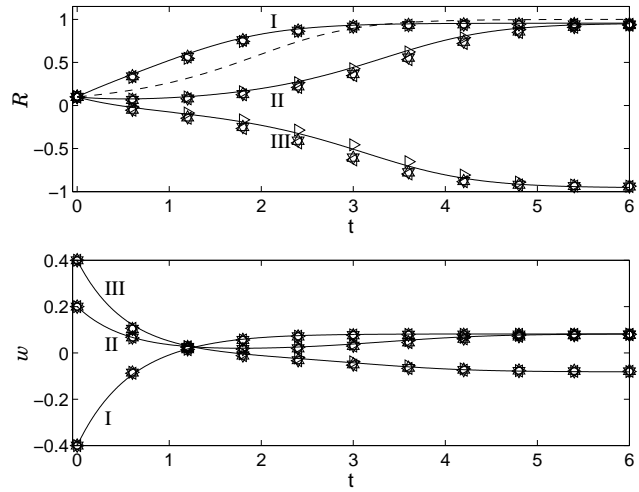,clip=,width=.7\textwidth}

\caption{The transient behavior predicted by Eq.\ (\ref{eq:first})
(solid line) is compared to that of the full system Eq.\
(\ref{eq:zj}) (triangles) and of its zeroth-order approximation
Eq.\ (\ref{eq:zeroth}) (dotted line) for $\sigma^2=0.5$
and  $K=3$. Three initial states are chosen,
having the same centroid's position $|Z|$, but different
configuration, and thus different $|W|$. Populations with
different size and frequency distribution are considered: $N=800$,
Gaussian distribution ($\bigtriangleup$); $N=800$ uniform
distribution ($\bigtriangledown$); $N=5$, uniform distribution
($\lhd$);  $N=2$ ($\rhd$). 
\label{fig:trans}}
\end{figure}

\begin{figure}[H]
\center
\epsfig{file=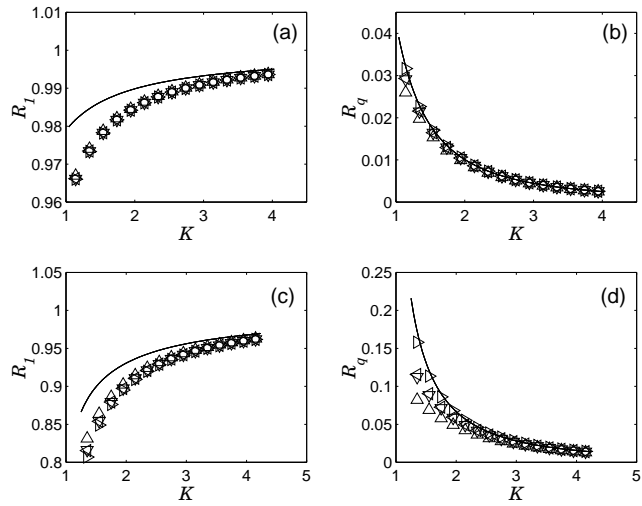,clip=,width=.7\textwidth}

\caption{The estimated values for the amplitude of the
centroid's oscillations (Eq.\ \ref{eq:zf}) and the quenching
radius (Eq.\ \ref{eq:rq}) versus the coupling constant $K$ (solid
lines) are compared to the numerically computed ones (triangles),
for $\sigma^2=0.2$ (cases {\it a} and {\it b}) and $\sigma^2=0.5$
(cases {\it c} and {\it d}). The accordance is higher for more
concentrated distributions, but is again almost unchanged when
populations of different size and frequency distribution are
considered (symbols as in Fig.\ \ref{fig:trans}). The accuracy
increases with the coupling, approaching at the meantime the
zeroth-order values $|Z|=1$ and $R_q=0$.\label{fig:pert} }  \end{figure}

{\it Comparison.} Let us now compare the behavior of the
equations for the macroscopic variables Eq.\ (\ref{eq:first})
with the numerical simulations of the original system described by
Eq.\ (\ref{eq:zj}). 
Fig.\ \ref{fig:trans} compares the
transient behavior of the full system (initialized in three
configurations having the same average $Z$, but different $W$) with the one
predicted according to Eq.\  (\ref{eq:first}). The transient behaviors
shown in Fig.\ \ref{fig:phase} are actually observed: the accuracy of the
approximation holds along the whole trajectory and remains nearly
unchanged for any population having the same $\sigma^2$ and $K$. Fig.\
\ref{fig:pert} compares the estimated and numerically computed values of the
asymptotic amplitude of the collective oscillations and the quenching radius.

We considered a
population of globally and strongly coupled oscillators with narrow
natural frequency distribution and showed how the dynamics can be reduced, via
an expansion in order parameters, to a low-order system containing all the
essential information (for the first-order expansion, two complex variables
instead of $N$ and two parameters instead of $N+1$). Qualitative as
well as quantitative results were then given for collective
properties of experimental relevance.  Although we restricted our
attention to simple oscillators, the presented method can in principle
be applied to other populations of globally and strongly coupled
elements with small parameter mismatch. Moreover, the macroscopic
description may furnish an instrument for studying collective
behaviors outside the phase locking region. In particular, the number
of the relevant terms in the expansion is expected to change as the
system approaches the bifurcation boundaries, linking the onset of
new regimes to a change in the dimensionality of the macroscopic
system.

The authors thank A. Bisgaard, S. Dan{\o}, A. Giorgilli, M. Granero, E.
Mosekilde, A. Porati, K. Sigmund, and P.G. S{\o}rensen for the fruitful
discussions. F. d'O. has been supported by TMR European grant
FMRX-CT96-0085. S. D. M. has been supported by a grant of the
Department of Mathematics, University of Milano.

%\begin{references}
 
[1] Y. Kuramoto, {\it
Chemical Oscillations, Waves and Turbulence}, Springer, Berlin
(1984).\\

[2] J. N. Blakely, D. J. Gauthier,  G. Johnson,  T. L. Carroll, and
L. M. Pecora, Chaos {\bf 10} 738 (2000). E. del Rio, J. R. 
Sanmartin, and O. Lopez-Rebollal, Int. J. Bif. and Chaos {\bf 8} 2255
(1998).\\

[3] C. M. Gray, P.  K\"{u}nig, A. K. Engel, and W. Singer, Nature
{\bf338}, 334 (1990).\\

[4] B. Lading, E. Mosekilde, S. Yanchuk, and Y. Maistrenko, Progr.
Theor. Phys. Suppl. {\bf139} 164 (2000).\\

[5] S. Dan{\o}, P. G. S{\o}rensen, and F. Hynne, Nature {\bf 402},
320 (1999).\\

[6] A. T. Winfree, J. Theoret. Biol. {\bf16} 15 (1967); {\it
Geometry of the Biological Time}, Springer, New York (1990).\\

[7] H. Haken, {\it Synergetics}, Springer-Verlag, Berlin, 1977. H.
Daido, Phys. Rev. Lett. {\bf 73}, 760 (1994); Int. J. Bif. and Chaos
{\bf 7} (1997).\\

[8] Y. Kuramoto, and I. Nishikawa, J. Stat. Phys. {\bf 49}, 569
(1987). L. L. Bonilla, J. C. Neu, and R. Spiegler, J. Stat. Phys. {\bf
67},313 (1992). P. C. Matthews, and S. Strogatz, Phys. Rev. Lett. {\bf
65}, 1701 (1990). J. D. Crawford, and K. T. R. Davies, Physica D {\bf
125}, 1 (1999). S. H. Strogatz, Physica D {\bf143}, 1 (2000).\\

[9] P. C. Matthews, R. E. Mirollo, and S. H. Strogatz, Physica D
{\bf 52}, 293 (1991).\\

[10] F. Hynne, P. G. S{\o}rensen, and K. Nielsen, J. Chem. Phys.
{\bf 92}, 1747 (1990).\\

$^\dag$  Calling $|z_j|=\rho_j$ and assuming a frame or reference
rotating at $\omega_0=\nsum{\omega_0}$,
$|Z-z_j|=\frac{1}{K}\left|(1-|z_j|^2 + i(\omega_j-\omega_0))z_j\right|
\leq{\frac{1}{K}\left[\left(1-|\rho_j|^2\right)\rho_j
+|\omega_j-\omega_0|\rho_j\right]}
\leq{(1+|\omega_j-\omega_0|)\rho_j\frac{1}{K}}
\leq(1+|\omega_j-\omega_0|)/K$.\\

%\end{references}

\end{document}